\documentclass[11pt,twoside]{article}
\usepackage{cozumel2005}
\usepackage{epsf}
\usepackage{psfig}
\usepackage{lscape}
\begin{document}
\title{Hi-Res Spectroscopy of a Volume-Limited Hipparcos Sample within 100 parsec}    
\author{P. L. Lim and J. A. Holtzman} 
\affil{Astronomy Department, MSC 4500, New Mexico State University, Box 30001, Las Cruces, New Mexico 88003, United States of America}    
\author{V. V. Smith}
\affil{National Optical Astronomy Observatory Gemini Science Center, Tucson, Arizona, United States of America}
\author{K. Cunha}
\affil{Department of Physics, University of Texas, El Paso, Texas, United States of America}

\begin{abstract} 
We obtain high-resolution echelle spectra of evolved Hipparcos stars within 100 parsec for detailed abundance analysis. We are particularly interested in subgiants because their ages can be derived accurately. Here, we present preliminary results for 10 subgiants and discuss the future work for this project. Our main scientific interest is to incorporate our results to constrain the star formation history in the solar neighborhood.
\end{abstract}


\section{Introduction}
Accurate parallax measurements of solar neighborhood stars from the Hipparcos catalog have enabled detailed studies of stellar properties including studies of the star formation history \citep{her00,ber01}. By looking at stellar abundances at different ages, we can constrain the star formation history of the solar neighborhood and study the history of chemical evolution. Subgiants are of particular interest because they lie in an area where isochrones are well separated, enabling more accurate age determination.

We have initiated a project to obtain high-resolution echelle spectra of evolved Hipparcos stars in a volume-limited sample within 100 parsec. We are performing detailed abundance analysis to investigate the chemical evolution and the star formation history of the solar neighborhood.

Currently, we have obtained spectra for about 50\% of our sample. We are attempting automated analysis to obtain fundamental attributes such as [Fe/H], effective temperature, surface gravity, and age. Here, we present initial results for 10 subgiants over a range of effective temperatures. We compare our results using several different methods to literature values, and discuss future development for this project.

\section{Sample}                      

\begin{figure}[!ht]
\plotone{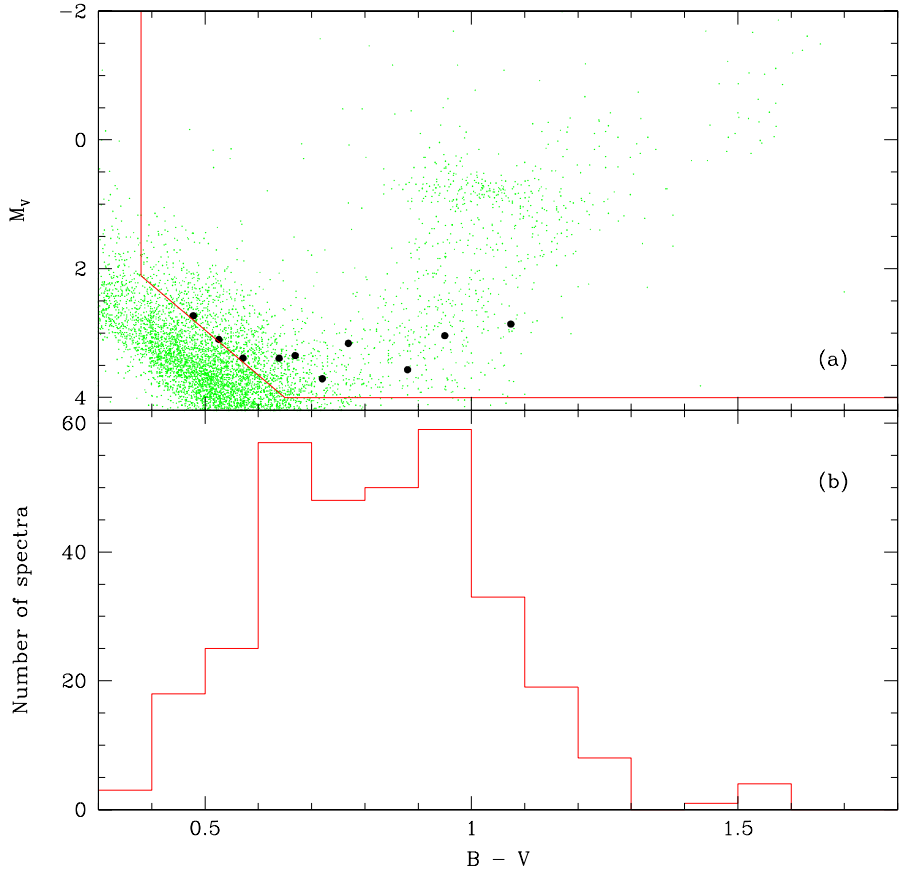}
\caption{(a) Color-magnitude diagram showing all Hipparcos stars within 100 parsec (small points) with sample cut lines (solid lines). Ten subgiants presented here are shown as large points. (b) Number of observed stars to date as a function of color. \label{fig1}}
\end{figure}

We select evolved Hipparcos stars with acceptable parallax uncertainties in the Northern Hemisphere within 100 parsec that have M$_{V} \leq$ 4 and B-V $\geq$ 0.38, as shown in Fig.~\ref{fig1}(a). To probe histories of star formation and chemical evolution, abundance analysis of subgiants is ideal because we can use the well-separated isochrones of the subgiant branch to obtain accurate ages. Therefore we start our analysis on subgiants first, of which the 10 that we are presenting here are also shown in Fig.~\ref{fig1}(a). They are chosen to cover a wide range of effective temperatures to test the accuracy of our automated method.

We are randomly selecting about 500 stars to observe as a representative sample. We exclude known binaries and multiple-star systems to avoid complications such as abundance ambiguity in spectroscopic binaries. We will also exclude stars with high rotational velocities and significant on-going mass loss.

By collecting our own data, we ensure a uniform sample, consistency in data reduction, and sufficient SNR to obtain abundances of various individual elements. We obtain high-resolution echelle spectra with the ARC 3.5-m telescope at the Apache Point Observatory, and use standard techniques for spectrum extraction and calibration. To date, we have obtained over 300 spectra, tallied in Fig.~\ref{fig1}(b), and are performing initial abundance analysis.

\section{Analysis}

\begin{figure}[!ht]
\plotone{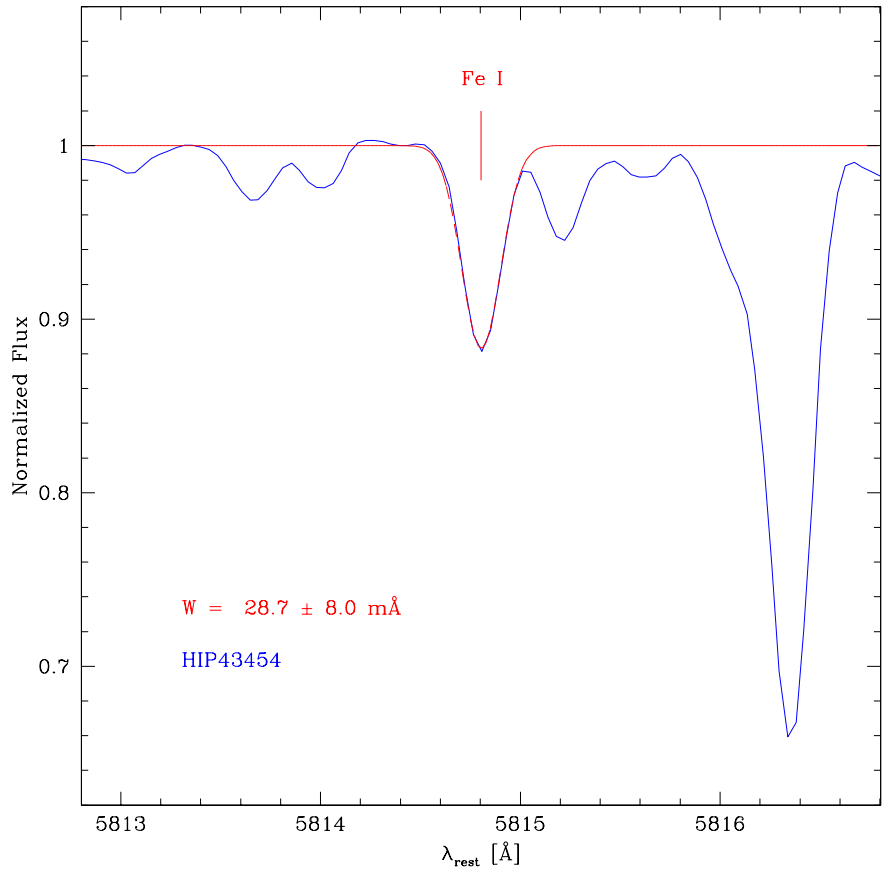}
\caption{An example of automatic nonlinear gaussian fitting for an Fe {\tiny I} line. The best fit is shown as solid line with flat continuum. The other solid line is the spectrum of HIP43454. Value of $W$ shown is the equivalent width of the fit. \label{fig2}}
\end{figure}

Due to the number of stars in our sample, we are attempting to automate the analysis wherever possible. We measure equivalent widths ($W$) of selected ``clean'' Fe {\tiny I} and Fe {\tiny II} lines using nonlinear least squares fitting with a gaussian profile. An example is shown in Fig.~\ref{fig2}. For the 10 stars discussed here, we also manually measure $W$ for comparison with the automatic results.

We use these measurements in MOOG \citep{sne02} with a grid of \citet{kur93} model atmospheres based on initial estimates of stellar properties. Our grid is interpolated from existing model atmospheres and has effective temperature ($T_{\mbox{eff}}$), logarithmic surface gravity ($\log g$), and mircoturbulence velocity ($\xi$) steps of 50 K, 0.1 cm s$^{-2}$, and 0.1 km s$^{-1}$ respectively. For each star, we obtain [Fe/H], $T_{\mbox{eff}}$, $\log g$, and $\xi$ by finding zero slopes of Fe {\tiny I} abundance versus excitation potential (EP) and reduced equivalent width (RW), and ionization balance between Fe {\tiny I} and Fe {\tiny II}. We can compare the $\log g$ obtained from parallax and $T_{\mbox{eff}}$ from color with those from spectroscopy.

With a derived metallicity, we fit Padova isochrones \citep{gir02} through a star to obtain its mass ($\mathcal{M}$), age ($\tau$), and ``trigonometric'' $\log g$. These isochrones are interpolated for sufficient fitting via closest-point match in the $L$,$T_{\mbox{eff}}$-plane and have [Fe/H] intervals of 0.05 dex. For the stellar luminosities, we apply corrections to the absolute magnitudes as suggested by \citet{lut73}. In future work, we will improve isochrone fitting by including probability weighting to allow for uncertainties in the photometry and distances, and overlap between isochrones.

\section{Initial Results - 10 Subgiants}

We present preliminary results for 10 subgiants. They are selected from the color-magnitude diagram to span a range of temperatures in order to evaluate the accuracy of our method over varying stellar properties (in this case, $T_{\mbox{eff}}$). We perform equivalent width measurement, abundance analysis, and isochrone fitting as described in the {\it Analysis} section. In Table~\ref{tab1}, we tabulate the results from the automated $W$ fitting.

\begin{table}[!ht]
\caption{Results for the 10 subgiants via automatic $W$ measurements. \label{tab1}}
\smallskip
\begin{center}
{\small
\begin{tabular}{lcccccc}
\tableline
\noalign{\smallskip}
HIP & $T_{\mbox{eff}}$ [K] & $\log g$ [cgs] & $\xi$ [km/s] & [Fe/H] & $\mathcal{M}$ [$\mathcal{M}_{\odot}$] & $\tau$ [Gyr] \\
\noalign{\smallskip}
 & $\pm$ 50 & $\pm$ 0.1 & $\pm$ 0.1 & & & \\
\noalign{\smallskip}
\tableline
\noalign{\smallskip}
13679 & 6100 & 3.5 & 2.0 & -0.08 $\pm$ 0.05 & 1.35 $\pm$ 0.22 & 3.2 $\pm$ 2.0\\
6869  & 6000 & 3.5 & 2.0 & -0.02 $\pm$ 0.09 & 1.22 $\pm$ 0.12 & 4.2 $\pm$ 1.5\\
9726  & 5950 & 3.9 & 1.8 & -0.15 $\pm$ 0.07 & 1.17 $\pm$ 0.08 & 5.1 $\pm$ 1.8\\
7276  & 5850 & 4.1 & 1.6 & +0.05 $\pm$ 0.05 & 1.21 $\pm$ 0.03 & 4.9 $\pm$ 0.4\\
43454 & 5750 & 3.8 & 1.4 & -0.01 $\pm$ 0.05 & 1.23 $\pm$ 0.06 & 4.8 $\pm$ 0.7\\
8159  & 5650 & 4.1 & 1.3 & +0.05 $\pm$ 0.04 & 1.12 $\pm$ 0.03 & 7.1 $\pm$ 0.5\\
39681 & 5250 & 3.5 & 0.9 & -0.43 $\pm$ 0.05 & 1.35 $\pm$ 0.69 & 2.8 $\pm >$ 5\\
52316 & 5150 & 3.9 & 1.0 & +0.01 $\pm$ 0.06 & 1.27 $\pm$ 0.05 & 4.4 $\pm$ 0.5\\
58080 & 5100 & 3.7 & 1.1 & +0.14 $\pm$ 0.07 & 1.55 $\pm$ 0.25 & 2.5 $\pm$ 1.7\\
61995 & 4800 & 3.5 & 0.9 & +0.26 $\pm$ 0.11 & 1.12 $\pm$ 0.39 & 8.0 $\pm >$ 5\\
\noalign{\smallskip}
\tableline
\end{tabular}
}
\end{center}
\end{table}

HIP13679 lies at the hook of main sequence turn-off, HIP39681 has considerable parallax uncertainty ($\frac{\sigma}{\pi_{o}} =$ 0.16), and HIP61995 turns out to be on the red giant branch. These cause substantial uncertainties in $\mathcal{M}$ and $\tau$ values. In the future, we will refine our selection criteria to minimize these issues.

\begin{figure}[!ht]
\plotone{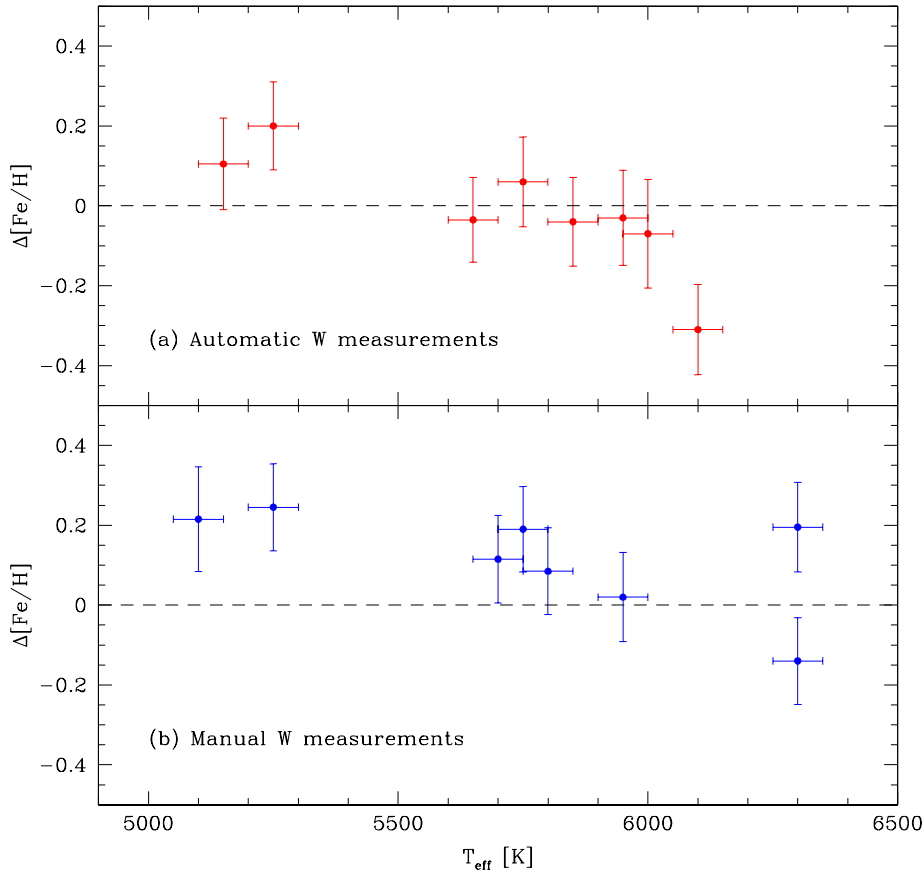}
\caption{Deviation of our [Fe/H] results from literature values for (a) automatic and (b) manual $W$ measurements. Error bars for [Fe/H] include both our uncertainties and assumed 0.1 dex errors for literature values. Error bars for $T_{\mbox{eff}}$ is from model atmosphere grid spacing of 50 K. HIP58080 and HIP61995 do not have published values for comparison. \label{fig3}}
\end{figure}

We compare our iron abundance from automated and manual $W$ measurements with literature values \citep{cay97,cay01,nor04} in Fig.~\ref{fig3}. On average, the automated measurements agree better with published values, except for HIP13679. More data on the lower and higher $T_{\mbox{eff}}$ ends are required to determine if there is any [Fe/H] trend with effective temperature.

\begin{figure}[!ht]
\plotone{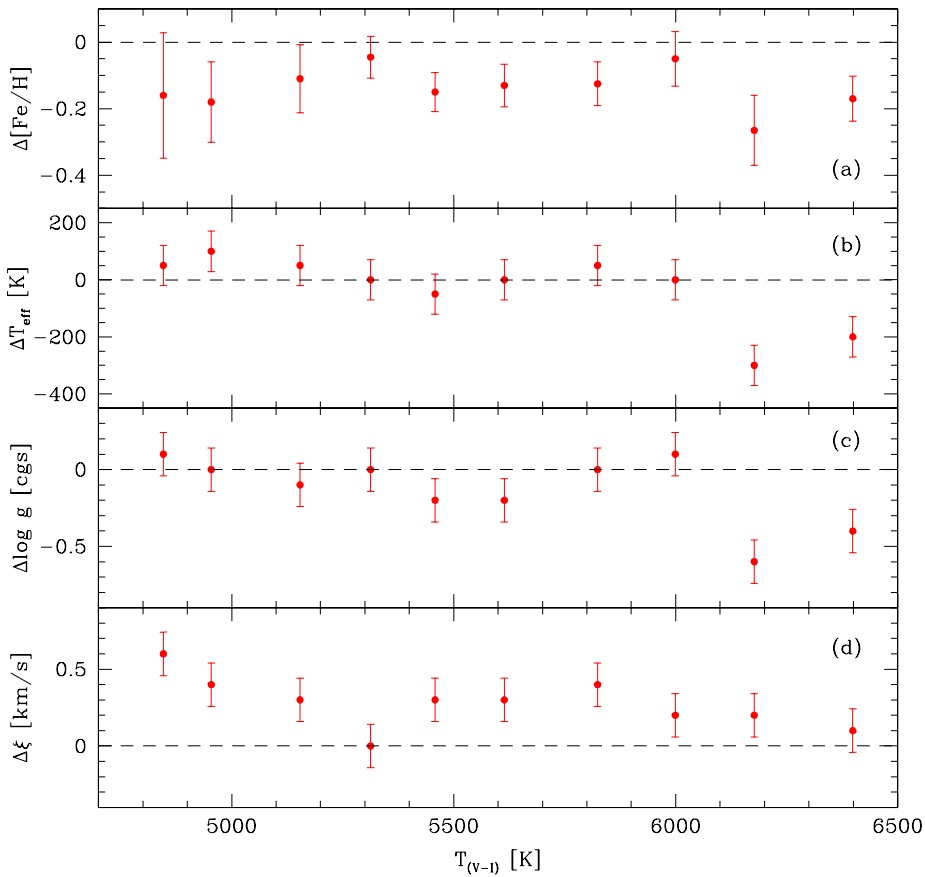}
\caption{Deviation of manual $W$ measurement results from the automatic for (a) metallicity, (b) effective temperature, (c) surface gravity, and (d) microturbulent velocity. Effective temperatures derived from $V-I$ for the 10 subgiants are plotted on the X-axis. \label{fig4}}
\end{figure}

We compare the derived stellar parameters from the automatic and manual $W$ measurements in Fig.~\ref{fig4}. We use $T_{\mbox{eff}}$ from $V-I$ \citep{bes98} for visual convenience. The ``automatic'' method underestimates [Fe/H] and overestimates $\xi$ compared to the ``manual.'' However the results for $\log g$ and $T_{\mbox{eff}}$ are comparable except at $>$6000$\mbox{\AA}$.

Two possible reasons for these deviations are systematic differences in the two $W$ measurement methods and slopes affected by outliers at the upper or lower limits of EP and RW during abundance analysis. This calls for more data and careful visual inspection of the results. For the former, we could fine-tune the automation, such as implementing dynamic continuum determination and line templates as a function of temperature. For the latter, we could add more lines near the EP and RW edges, or formulate an analysis method less dependent of the slopes.

\begin{figure}[!ht]
\plotone{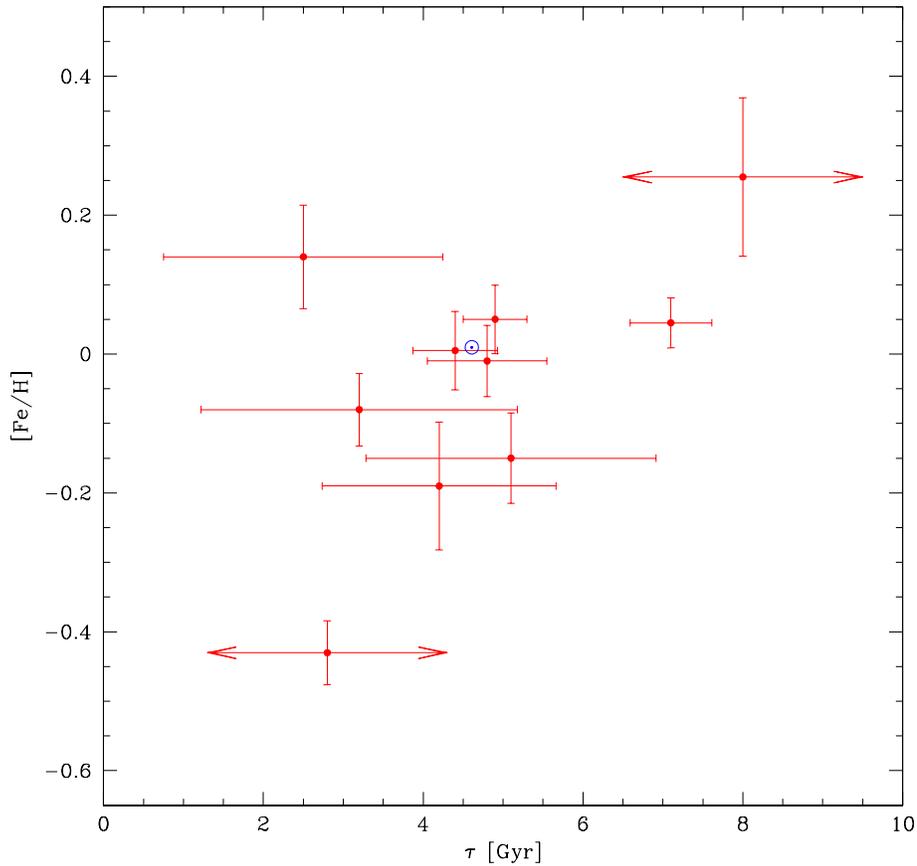}
\caption{Age-metallicity relationship based on the 10 subgiants (solid points). Age uncertainty larger than 5 Gyr is shown as an arrow. Sun is shown as solar symbol for reference. \label{fig5}}
\end{figure}

From our initial results of 10 stars from automated $W$ fitting, we show the age-metallicity relationship (AMR) in Fig.~\ref{fig5}. \citet{edv93} reported a real scatter of metallicity with respect to age, and much published literature states no tight correlation in AMR for [Fe/H]. Our present results here show no surprising contradictions.

\section{Future Work}

We will continue to obtain data until our targeted sample size of $\sim$500 stars is achieved. We will improve our current automation algorithm for more robust $W$ measurements and better accuracy in derived stellar properties. In addition to Fe, we will extend abundance analysis to  C, O, Mg, and many other elements made possible by high-resolution spectroscopy.

Once we have sufficient abundance information, we will be able to incorporate this into a study of the star formation history of the solar neighborhood, using observed values to constrain the models. We will also investigate the abundance distribution of heavy elements and the age-metallicity relation. 

By selecting a volume-limited sample, we are biased towards thin-disk stars that have smaller velocity dispersion from the plane. We are also biased towards more metal-rich stars due to a lower-limit cutoff in the color. In the sample of subgiants, we are biased towards less massive stars as they have a longer timescale in that evolutionary stage than their more massive counterparts. We will attempt to correct for these biases in our analysis.

A useful by-product of this project will be a database of high resolution spectra of volume-limited sample of solar neighborhood stars, yielding a catalog of fundamental stellar properties, kinematics, and detailed abundances.

\acknowledgements             
Support for this work is provided by the National Science Foundation. We also thank the organizing committee of Cozumel Resolved Stellar Populations Meeting 2005 for the wonderful time in Canc\'un, M\'exico.


\end{document}